\documentstyle[12pt,axodraw]{article}

\newcommand{\eq}[1]{eq.(\ref{#1})}

\def\ghost#1{\vrule height#1 depth#1 width0pt \displaystyle}
\def\const{\mbox{const}}
\def\e{\mbox{e}}
\def\ch{\mbox{ch}}

\def\sh{\mbox{sh}}

\def\half{{1 \over 2}}
\begin{document}

\title{What Becomes of Vortices in Theories with Flat Directions}
\author{
  A.A.Penin, V.A.Rubakov, P.G.Tinyakov and S.V.Troitsky\\ 
  {\small {\em Institute for Nuclear Research of the 
  Russian Academy of Sciences,}}\\
  {\small {\em 60th October Anniversary 
  prospect, 7a, Moscow 117312, Russia.}}
  }
\date{August 1996}
\maketitle
\begin{abstract}
In many theories with flat directions of scalar potential, static
vortex solutions do not exist for a generic choice of vacuum. In two
Euclidean dimensions, we find their substitutes --- constrained
instantons consisting of compact core formed by
Abrikosov--Nielsen--Olesen vortex and long-ranged cloud of modulus
field. In (3+1) dimensions, an initial compact configuration of string
topology evolves in such a way that at every point in space the
modulus relaxes, in a universal manner, to one and the same value
characteristic to the theory.
\end{abstract}

\newpage

{\bf 1.} Flat directions of scalar potential determining moduli spaces
of physically inequivalent vacua are inherent in many supersymmetric
gauge theories. Some of these theories exhibit topological properties
allowing for Abrikosov--Nielsen--Olesen vortices. However, at least in
several examples, static vortex solutions actually do not exist for
generic choice of vacuum \cite{Witten}. The purpose of this paper is
to analyse this situation in some detail.

This analysis may be of interest from two points of view.  First,
vortices are instantons in two dimensions. In the absence of exact
instanton solutions, the instanton transitions are dominated by
constrained instantons \cite{Affleck}. Hence, our first purpose is to
describe constrained instantons in two-dimensional theories with flat
directions. We shall see that, unlike the situation in
Yang--Mills--Higgs theories in four dimensions, the constraint is
required to stabilize the instanton against {\em extending} to
infinite size. This and other peculiarities of the constrained
instantons in two dimensions, which we describe below, are obviously
due to the logarithmic behavior of classical massless moduli fields
at large distances.

The second aspect is that vortices are cosmic strings in four
dimensions. Since the static vortex solutions do not exist in models
under discussion, an initial configuration with vortex topology
(created, say, by the Kibble mechanism) will evolve in time. We shall
see that this evolution is rather peculiar: i) at given point in space
and at sufficiently large times the evolution is slow (logarithmic in
time) and independent of the details of the initial configuration; ii)
the evolution (slowly) drives the moduli field to a certain value
everywhere in space; in other words, the presence of the string
results in slow transition from the original vacuum to a particular
one which is fixed in a given theory.

{\bf 2.} To be specific, let us consider $U(1)$ gauge theory with two
complex scalars of opposite charges, described by the following
Lagrangian,
\begin{equation}
L = - {1\over 4} F_{\mu\nu}^2 + |D_{\mu}\phi_1|^2 + 
|D_{\mu}\phi_2|^2 - V(\phi_1,\phi_2),
\label{4*}
\end{equation}
where 
\[
V(\phi_1,\phi_2) = 
{\lambda\over 2} (|\phi_1|^2 - |\phi_2|^2 -\eta^2)^2 .
\]
This theory at $\lambda=e^2$ ($e$ is the gauge coupling) may be viewed
as the bosonic part of $N=1$, $d=4$ or $N=2$, $d=2$ supersymmetric
theory with zero superpotential and with Fayet--Illiopoulos $D$-term 
proportional to $\eta^2$. The moduli space, up to 
gauge and global transformations, is parameterized by one parameter
$u$, 
\[
\phi_1=\mbox{real}=v_1=\eta\ch u ,
\]
\begin{equation}
\phi_2=\mbox{real}=v_2=\eta\sh u .
\label{5*}
\end{equation}
The gauge symmetry is broken for any choice of vacuum, the mass of the
gauge boson being
\[
m_V = e\eta \sqrt{2\ch 2u}.
\]
There exists one massless Goldstone field corresponding to broken global
symmetry $\phi_1,\phi_2 \to \e^{i\alpha}\phi_1,\e^{i\alpha}\phi_2$,
and massless modulus. The latter will be of primary interest for our
purposes. The modulus field
with canonical kinetic term is related to the
variable $u$ as follows, 
\begin{equation}
z = \eta \int\limits_0^u du \sqrt{2\ch 2u}, 
\label{5**}
\end{equation} 
so that $dz/\sqrt{2}$ is the length element along the moduli curve 
(\ref{5*}).

In a given vacuum, the static $O(2)$-symmetric configurations with
vortex topology have the following structure,
\[
\begin{array}{rcl}
\phi_1 & = & f_1(r)\e^{i\theta} ,\\
\phi_2 & = & f_2(r)\e^{-i\theta} ,\\
A_{\alpha} & = & \ghost{12pt} -{1\over e} \epsilon_{\alpha\beta}
a(r) {x_{\beta}\over r}, ~~~~\alpha,\beta = 1,2 \\
\end{array}
\]
with $f_1(0)=f_2(0)=a(0)=0$ and 
\begin{equation}
\begin{array}{rcl}
f_1(r\to\infty) & = & v_1 ,\\
f_2(r\to\infty) & = & v_2 ,\\
a(r\to\infty) & \to & 1/r .\\
\end{array}
\label{6**}
\end{equation}

The reason for the absence of the static vortex solution at $v_2\neq
0$, i.e. at $z_0\neq 0$, where $z_0$ is the value of the modulus field
in the prescribed 
vacuum, is as follows. The modulus field extends to large
distances. At these distances other fields of vortex-like
solutions die away, and the modulus field becomes free. The solution
to free massless equation logarithmically increases with $r$, so the
modulus field grows to infinity instead of approaching its vacuum
value $z_0$. Hence, the boundary conditions (\ref{6**}) cannot be
satisfied. This argument does not work for a special choice of vacuum,
\begin{equation}
\begin{array}{l}
\phi_1^{(0)} = \eta,\\
\phi_2^{(0)} = 0.\\
\end{array}
\label{6*}
\end{equation}
In this special case, the field $\phi_2$, and hence the modulus field,
can be set to zero everywhere in space, and the vortex solution 
exists and coincides with the standard one 
appearing in the theory with one Higgs field.

{\bf 3.} Let us consider first the theory (\ref{4*}) in two Euclidean
dimensions. To obtain the constrained instanton in a general vacuum,
let us impose the constraint that the fields $\phi_1$ and $\phi_2$ are
equal, up to gauge transformation, to their vacuum values
(\ref{5*}) at large but
finite distance $r=R$, i.e.
\begin{equation}
\begin{array}{cc}
f_1(R)=v_1 & \\
f_2(R)=v_2, & \ghost{12pt} R\gg{1\over e\eta} 			
\end{array}
\label{7*}
\end{equation}
Outside this circle the fields remain constant.  It is straightforward
to solve the field equations numerically; the results for $f_1(r)$ and
$f_2(r)$ are presented in fig.1. 
At large enough $R$, the solution has the core of the size
\[
r_0 \sim 1/e\eta
\]
and the cloud extending to $r=R$.
For very large $R$, the function
$f_1$ reaches the value $f_1=\eta$ just outside the core, while the
function $f_2$ remains small in this region. The gauge field
configuration $a(r)$ coincides with the gauge field of the vortex in
the model with one Higgs field with expectation value $\eta$. In other
words, the field configuration inside and just outside the core is
the same for all vacua and coincides with the vortex solution existing in the special
vacuum (\ref{6*}) (up to corrections of order $1/\ln R$). Outside the
core, the fields $f_1$ and $f_2$ logarithmically tend to their values
(\ref{7*}) in such a way that $f_1^2 - f_2^2  = \eta^2$, i.e., only
the modulus field varies. 

This behavior is straightforward to understand analytically.  Let us
assume for a moment that the structure of the core is the same for all
choices of vacua and coincides with the core structure of the vortex
solution existing in the special vacuum (\ref{6*}). In particular,
up to corrections of order $1/\ln R$,
\begin{equation}
\begin{array}{r}
f_1= \eta ,\\
f_2=0\\ 			
\end{array}
\label{8*}
\end{equation}
just outside the core. The relations\footnote{One can show that these
relations are indeed consistent with the field equations at $r\gg
1/e\eta$.}  $f_1^2 - f_2^2 =\eta^2 $ and $a(r)=1/r$ are required
well outside the core
for the action not to increase with $R$. The remaining 
modulus field $z(r)$
obeys the free massless equation
\[
z''+ {1\over r} z' =0
\]
whose solution satisfying the boundary conditions (\ref{7*}), (\ref{8*}) is,
again up to corrections suppressed by $1/\ln R$, 
\begin{equation}
z(r) = z_0 {\ln r/r_0 \over \ln R/r_0},
\label{9*}
\end{equation}
where 
$z_0$ is the value of the modulus in the prescribed vacuum. 
The fields $f_1(r)$ and $f_2(r)$ outside
the core are then equal to $f_1(r)=\eta\ch u(r)$, $f_2(r)=\eta\sh
u(r)$, where $u(r)$ and $z(r)$ are related by \eq{5**}. Therefore, the
fields $f_1$ and $f_2$ approach their boundary values logarithmically. 

To understand why the structure of the core is universal (i.e., does
not depend on the choice of the vacuum), let us calculate the part of
the action which comes from the region outside the core, 
\[
S_{\mbox{\tiny outside}} = \int\limits_{r_0<r<R} d^2x ~
\half (\partial_{\mu}z)^2 = {\pi z_0^2 \over \ln R/r_0}. 
\]
This part is small at large $R$. In other words, it costs essentially 
no action to develop the modulus field far outside the core.
Hence the core is free to settle down in such a way that
its own action is minimized, with no condition imposed on the 
modulus field just outside the core. This minimum is clearly 
universal and coincides with the solution in the vacuum (\ref{6*}).

This argument is valid up to corrections of order $(\ln{R/r_0})^{-1}$.
Hence, the corrections to the action of this order,
coming from the core region, are not yet ruled out.
To see that there are no such corrections, let us make use
of the following scaling argument. The constrained 
instanton solutions corresponding to two sets of parameters 
$(z_0,~R)$ and $(z_0',~R')$ coincide at $r<R$, $r<R'$ provided
that (see eq.~(\ref{9*}))
\begin{equation}
{z_0\over\ln{R/r_0}}= {z_0'\over\ln{R'/r_0}}\ .
\end{equation}
So, the contribution of the core region to the action, $S_{\mbox{\tiny
core}}$,
depends only on the combination $z_0(\ln{R/r_0})^{-1}$.
At small values of this parameter, the field $z(r)$, and therefore, 
the function
$f_2(r)$, is small inside the core, and its contribution to 
$S_{\mbox{\tiny core}}$
is quadratic in its amplitude. Hence
\begin{equation}
S_{\mbox{\tiny core}}=
S_0(\eta) + O\left({z_0^2\over\ln^2{R/r_0}}\right)\ ,
\end{equation}
where $S_0(\eta)$ is the action for the vortex solution
in the special vacuum (\ref{6*}) (say, $S_0=\pi\eta^2$ at 
$\lambda =e^2$ \cite{Bogomol}). We conclude that the action 
of the constrained instanton at large $R$ is
\begin{equation}
S=S_0(\eta) + {\pi z_0^2\over\ln{R/r_0}} + 
O\left({1\over\ln^2{R/r_0}}\right)\ .
\label{11*}
\end{equation}
A somewhat surprising feature of this expression is 
that at sufficiently large $R$, the action
is proportional to $\eta^2$ and not to the vacuum expectation
values of the fields $\phi_1$ or $\phi_2$. This means that the 
instanton contribution to the path integral is sizeable
even at large $v_1$ and $v_2$ provided that $\eta$ is not too large.

To summarize, the constrained instanton consists of the universal
(independent of the  vacuum at $r=\infty$) 
core of size $r_0\sim 1/e\eta$
that dominates  the instanton action, and the long--ranged cloud 
of the modulus field determining  the correction to the action 
of order  $(\ln{R/r_0})^{-1}$.

{\bf 4}. Let us now turn to $(3+1)$ dimensions. Suppose that initially a
compact configuration of string topology exists in a generic vacuum
of the model (\ref{4*}). As is clear from the above discussion, the
configuration will evolve in such a way that the size of its massless
cloud, $R$, will increase 
and the vacuum (\ref{6*}) will finally settle down
everywhere in space, with the usual Abrikosov--Nielsen--Olesen string
remaining near $r=0$. Indeed, the latter evolution reduces the static
energy per unit length, eq.(\ref{11*}).

Our purpose is to evaluate the behavior of the fields during this
process.  Clearly, the fields near the light cone depend on the
details of the initial configuration. On the other hand, the behavior
of the fields at large times and deep inside the light cone is
universal. To show this, let us first note that outside the core, the only
varying field is the modulus $z$. It obeys the free massless equation,
\begin{equation}
- \partial^2_t z + z'' + \frac{1}{r} z' =0      
\label{13*}
\end{equation}
(we still consider $O(2)$-symmetric configurations; this 
restriction in fact does
not lead to loss of generality, as only $s$-waves survive at large
$r$). Let us consider an initial configuration $z=z_i(r)$ that
vanishes at $r=r_0$ where $r_0$ is of order of the core radius, and
approaches the vacuum value at some $r=r_1>r_0$ (but still $r_1$ is of
order $r_0$),
\begin{equation}
z_i(r)= z_0,~~~~~r>r_1\ .
\label{13++}
\end{equation}
Let us assume for simplicity that initially
\begin{equation}
\partial_t z(t=0) =0
\label{13**}
\end{equation}
(this assumption is unimportant and may be safely 
dropped). Let us note also
that the modulus field is zero in the core at all times (otherwise the
energy of the core would become larger than $S_0$), so that we have a
condition
\begin{equation}
z(r_0) = 0
\label{13+}
\end{equation}
at any $t$. These conditions determine the solution to eq.(\ref{13*}) 
uniquely.

The complete set of solutions to eq.(\ref{13*}) with the conditions 
(\ref{13**}), (\ref{13+}) is 
\[
\cos (\omega t) Z_{\omega}(r)\ ,
\]
where $\omega$ is a continuous parameter,
\[
Z_{\omega}(r) = N_{\omega} \left[ J_0(\omega r_0)  N_0(\omega r)
                -  N_0(\omega r_0) J_0(\omega r) \right]\ ,
\]
$J_0$ and $N_0$ are the Bessel functions and
\[
N_{\omega} =
\left[ \frac{\omega}{J_0^2(\omega r_0) +
N_0^2(\omega r_0)} \right]^{1/2}\ .
\]
The functions $Z_{\omega}(r)$ are normalized (with measure $r\,dr$) to
$\delta (\omega - \omega')$. Therefore, the solution to the problem
(\ref{13*}), (\ref{13++}), (\ref{13**}), (\ref{13+}) is
\begin{equation}
z(r,t) = \int~d\omega~ z_{\omega} \cos(\omega t) Z_{\omega}(r)
\label{14*}
\end{equation}
with
\[
z_{\omega} =   \int~rdr~ z_i(r) Z_{\omega}(r)\ .
\]
At large $t$  and $r $, and at $(t-r) \sim t$ 
(deep inside the light cone and
 far from the core) the integral (\ref{14*}) is dominated by small 
$\omega$, namely, 
$\omega \sim t^{-1}$. It is straightforward to see that at 
small $\omega$, the
function $z_{\omega}$ is independent of the details of the initial
configuration and is equal to
\[
z_{\omega} =          \frac{z_0}{\omega^{3/2} \ln (\omega r_0)}\ .
\]
So, in this regime
\begin{equation}
z(r,t) = \frac{\pi}{2} z_0
\int~\frac{d\omega}{\omega \ln^2(\omega r_0)}
~\cos(\omega t) \left[ N_0(\omega r) - \frac{2}{\pi} \ln (\omega r_0)
J_0 (\omega r)\right]\ .
\label{15+}
\end{equation}
Clearly, $z(r,t)$ depends on $r$ and $t$ logarithmically, and at 
$r_0\ll r \ll t$ the solution has a particularly simple form
\begin{equation}
 z(r,t) = z_0 \frac{\ln r/r_0}{\ln t/r_0}
\label{15*}
\end{equation}
up to corrections suppressed by $(\ln{r/r_0})^{-1}$,
$(\ln{t/r_0})^{-1}$. Note that eq.(\ref{15*}) coincides with the
constrained instanton solution (\ref{9*})
whose radius $R$ linearly increases with time.

The solution deep inside the light cone, eq.(\ref{15+}), does not
depend on the details of the initial configuration, i.e., it is
universal.  At a given point $r$ in space, the modulus field relaxes
to zero logarithmically in time, eq.(\ref{15*}), i.e., the vacuum
(\ref{6*}) logarithmically settles down.  A typical evolution of the
field $f_2$ obtained by solving numerically the exact non-linear field
equations coming from the Lagrangian (\ref{4*}), is presented in
fig.2. Fig.3 demonstrates that the numerical solutions are in agreement
with \eq{15*}. 

The peculiar properties of string-like objects in the model (\ref{4*})
are entirely due to the presence of the modulus field in the theory.
So, these properties should be generic in many models with flat
directions. On the other hand, these features occur because of the specific
behavior of massless classical fields in two dimensions, so they do
not show up in the case of point-like objects such as monopoles.

\section*{Acknowledgments}

\noindent
The authors are indebted to A.N. Kuznetsov and M.V. Libanov for
helpful discussions. The work of A.P. is supported in part by
INTAS grant 93-2492-ext (research program of 
International Center for Fundamental Physics in Moscow). The work of
V.R. and S.T. is supported in part by Russian Foundation for Basic
Research grant 96-02-17449a and INTAS grant 94-2352. The work of
P.T. is supported in part by Russian Foundation for Basic Research
grant 96-02-17804a and INTAS grant 94-2352. The work of S.T. is 
supported in part by Soros fellowship for graduate students.

\newpage
\section*{Figure captions}
\noindent
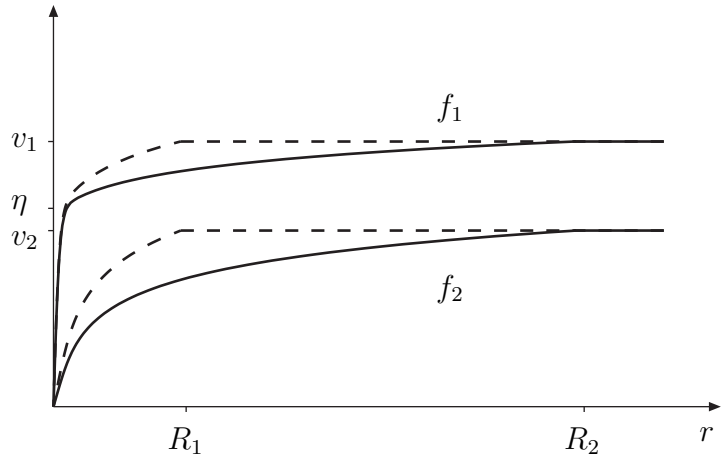
{\bf Figure 1.}  The functions $f_1(r)$ and $f_2(r)$ of constrained
instanton at $\lambda =e^2=1$, $\eta =1$, $v_1=1.34$, $v_2=0.89$, for
two different values of the constraint parameter $R_1=25$ (dashed
curves) and
$R_2=100$ (solid curves).
\vspace{0.5cm} 

\noindent
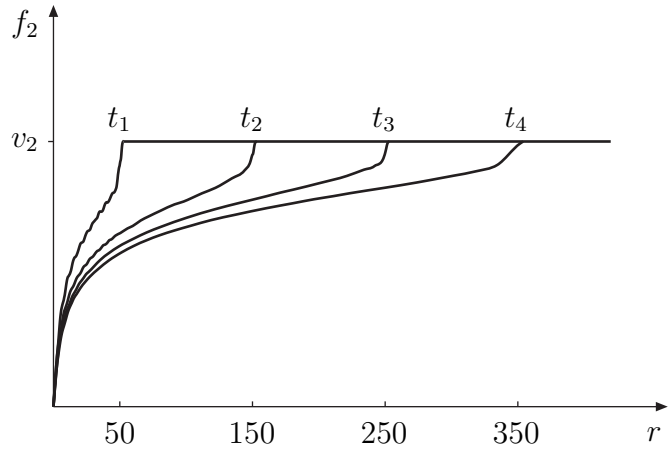
{\bf Figure 2.} The component $f_2(r)$ of a numerical solution to the
exact field equations at different times, $t_1=50$, $t_2=150$,
$t_3=250$, $t_4=350$. The function $f_1(r,t)$ for this solution obeys
$f_1^2-f_2^2=\eta^2$ to better than 0.5 per cent
outside the core. The parameters are the same as
in fig.1. The initial configuration had the size of order one.
\vspace{0.5cm} 

\noindent
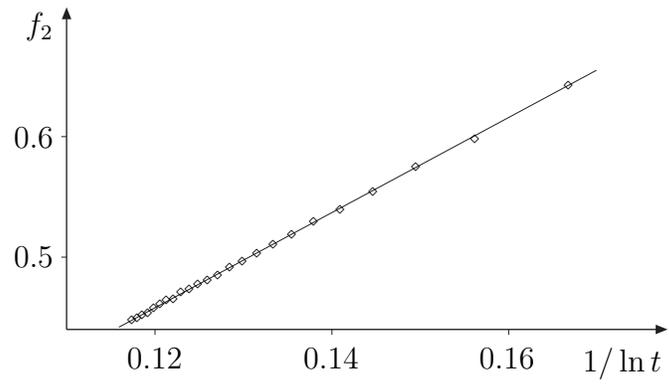
{\bf Figure 3.} The value of $f_2$ at $r=200$ as function of $1/\ln t$
for the exact solution shown in fig.2. Solid line is a linear fit 
$f_2 = v_2 \ln (r/r_0) / \ln (t/r_0)$ with fitted value of $r_0$ equal
to 2.39. Note that at small $z(r,t)$ one has $f_2(r,t) = \const\cdot
z(r,t)$.

\newpage
\vspace*{2cm}

\begin{center}
\begin{picture}(250,150)(0,0)

\LongArrow(0,0)(250,0)
\LongArrow(0,0)(0,150)

\Line(-2,100)(0,100)
\Line(-2,74.8)(0,74.8)
\Line(-2,66.4)(0,66.4)

\Text(-16,100)[l]{$v_1$}
\Text(-16,76)[l]{$\eta$}
\Text(-16,63)[l]{$v_2$}

\Text(250,-8)[tr]{$r$}

\Line(50,-2)(50,0)
\Text(50,-8)[t]{$R_1$}
\Line(200,-2)(200,0)
\Text(200,-8)[t]{$R_2$}

\Text(150,110)[b]{$f_1$}
\Text(150,50)[t]{$f_2$}
\SetWidth{1.0}
\Curve{(0, 0 )
(0.5172, 16.4361 )(1.16815, 35.1859 )
(1.98743, 52.8251 )(3.01858, 65.6325 )
(4.31639, 72.7881 )(5.9498, 76.2028 )
(8.00562, 78.0168 )(10.5931, 79.413 )
(13.8496, 80.8021 )(17.9484, 82.267 )
(23.107, 83.8087 )(29.5997, 85.4177 )
(37.7714, 87.0863 )(48.0562, 88.8085 )
(61.0008, 90.5793 )(77.2928, 92.3945 )
(97.7979, 94.2504 )(123.606, 96.144 )
(156.087, 98.0722 )(196.969, 100.033 )}
\Line(196,100)(230,100)

\Curve{(0, 0 )
(0.5172, 1.80307 )(1.16815, 4.0626 )
(1.98743, 6.87322 )(3.01858, 10.2753 )
(4.31639, 14.1826 )(5.9498, 18.372 )
(8.00562, 22.585 )(10.5931, 26.6541 )
(13.8496, 30.5322 )(17.9484, 34.2372 )
(23.107, 37.801 )(29.5997, 41.25 )
(37.7714, 44.605 )(48.0562, 47.8815 )
(61.0008, 51.0915 )(77.2928, 54.245 )
(97.7979, 57.3492 )(123.606, 60.4108 )
(156.087, 63.4347 )(196.969, 66.4253 )}
\Line(196,66.4253)(230,66.4253)

\DashCurve{(0, 0 )
(0.349135, 11.5719 )(0.759494, 24.5639 )
(1.24154, 38.0666 )(1.8074, 50.7372 )
(2.47205, 61.2652 )(3.25325, 68.9896 )
(4.1703, 74.119 )(5.24816, 77.4059 )
(6.5143, 79.6639 )(8.00062, 81.4826 )
(9.7464, 83.1825 )(11.7983, 84.8945 )
(14.2087, 86.6536 )(17.0382, 88.461 )
(20.3639, 90.3106 )(24.2679, 92.1969 )
(28.8534, 94.1153 )(34.243, 96.0624 )
(40.5699, 98.0357 )(48.0062, 100.033 )}{5}
\DashLine(48,100)(230,100){5}

\DashCurve{(0, 0 )
(0.349135, 1.80337 )(0.759494, 3.91526 )
(1.24154, 6.38115 )(1.8074, 9.23867 )
(2.47205, 12.5026 )(3.25325, 16.1458 )
(4.1703, 20.0869 )(5.24816, 24.1993 )
(6.5143, 28.3438 )(8.00062, 32.4086 )
(9.7464, 36.3317 )(11.7983, 40.0978 )
(14.2087, 43.7182 )(17.0382, 47.2132 )
(20.3639, 50.6019 )(24.2679, 53.9007 )
(28.8534, 57.1227 )(34.243, 60.2785 )
(40.5699, 63.3767 )(48.0062, 66.4253 )}{5}
\DashLine(48,66.4253)(230,66.4253){5}
\Text(150,-70)[]{Fig.1}
\end{picture}
\end{center}

\newpage
\vspace*{2cm}

\begin{center}
\begin{picture}(250,150)(0,0)

\LongArrow(0,0)(230,0)
\LongArrow(0,0)(0,150)
\Text(-16,150)[lt]{$f_2$}
\Line(-2,100)(0,100)
\Text(-16,100)[l]{$v_2$}
\Text(230,-8)[tr]{$r$}

\Line(25,0)(25,2)
\Text(25,-6)[t]{50}
\Text(25,106)[b]{$t_1$}

\Line(75,0)(75,2)
\Text(75,-6)[t]{150}
\Text(75,106)[b]{$t_2$}

\Line(125,0)(125,2)
\Text(125,-6)[t]{250}
\Text(125,106)[b]{$t_3$}

\Line(175,0)(175,2)
\Text(175,-6)[t]{350}
\Text(175,106)[b]{$t_4$}

\SetWidth{1.0}

\Curve{(0, 0 )
(0.15, 2.42005 )(0.300005, 4.86599 )
(0.450005, 7.50225 )(0.60001, 9.60248 )
(0.75001, 11.9178 )(0.900015, 13.6836 )
(1.05001, 15.4662 )(1.20002, 17.0023 )
(1.35002, 18.8739 )(1.50002, 20.3975 )
(1.65003, 22.134 )(1.80002, 24.0023 )
(1.95003, 25.9943 )(2.10003, 27.8997 )
(2.25004, 29.8333 )(2.40003, 31.6892 )
(2.55004, 33.348 )(2.70004, 34.7173 )
(2.85004, 35.8817 )(3.00005, 36.8671 )
(3.15004, 37.4966 )(3.30005, 38.1768 )
(3.45005, 38.7523 )(3.60005, 39.3018 )
(3.75006, 39.7781 )(3.90006, 40.3964 )
(4.05006, 41.1801 )(4.20007, 41.8637 )
(4.35006, 42.7691 )(4.50007, 43.812 )
(4.65007, 44.8345 )(4.80007, 45.9099 )
(4.95008, 46.8671 )(5.25008, 48.3785 )
(5.85008, 49.4865 )(6.6001, 51.3232 )
(6.9001, 52.6813 )(7.2001, 54.1487 )
(7.65012, 55.6407 )(8.70013, 57.0923 )
(9.45014, 59.2218 )(10.0502, 61.1701 )
(11.4002, 62.5856 )(12.0002, 64.402 )
(12.6002, 65.8176 )(13.9502, 67.1307 )
(14.4002, 68.4809 )(15.7502, 70.384 )
(16.6502, 71.643 )(16.9503, 72.6498 )
(17.2503, 73.3672 )(17.5503, 73.6216 )
(17.8503, 73.5506 )(18.1503, 73.571 )
(18.4503, 73.7849 )(18.7503, 74.3728 )
(19.0503, 75.2838 )(19.2003, 75.7803 )
(19.8003, 76.9471 )(20.5503, 77.1036 )
(21.3003, 78.4528 )(22.0503, 80.2331 )
(23.1003, 81.1171 )(23.5504, 82.1735 )
(23.7004, 82.6161 )(23.8504, 83.4347 )
(24.0004, 84.6036 )(24.1504, 85.8862 )
(24.3004, 87.4099 )(24.4504, 88.4144 )
(24.6004, 89.5 )(24.7504, 90.2883 )
(24.9004, 90.7782 )(25.0504, 91.7173 )
(25.2004, 92.5349 )(25.3504, 93.0608 )
(25.5004, 94.1824 )(25.6504, 96.0518 )
(25.8004, 97.8829 )(25.9504, 99.0889 )
(26.1004, 99.6768 )(26.4004, 100.012 )}
\Line(26.4004, 100.012 )(210, 100.012 )

\Curve{(0, 0 )
(0.15, 2.29955 )(0.300005, 4.65991 )
(0.450005, 6.68243 )(0.60001, 9.01352 )
(0.75001, 11.2827 )(0.900015, 13.5338 )
(1.05001, 15.5609 )(1.20002, 17.4696 )
(1.35002, 19.3603 )(1.50002, 20.795 )
(1.65003, 22.0811 )(1.80002, 23.3795 )
(1.95003, 24.3074 )(2.25004, 25.9775 )
(2.55004, 27.5484 )(2.85004, 29.1126 )
(3.15004, 30.9313 )(3.30005, 32.0135 )
(3.60005, 33.8457 )(3.90006, 35.3548 )
(4.20007, 36.4538 )(4.50007, 37.2466 )
(4.80007, 38.0957 )(5.10008, 38.9988 )
(5.40008, 40.0788 )(5.70008, 41.2477 )
(6.00009, 42.3851 )(6.4501, 43.6859 )
(6.9001, 44.5225 )(7.35011, 45.1936 )
(7.80011, 46.1216 )(8.25013, 47.2996 )
(8.70013, 48.3952 )(9.15014, 49.1137 )
(9.60014, 49.6228 )(10.0502, 50.1734 )
(10.5002, 50.9775 )(10.9502, 52.0034 )
(11.4002, 52.7883 )(11.8502, 53.1858 )
(12.3002, 53.491 )(12.7502, 54.0879 )
(13.2002, 54.9606 )(13.6502, 55.6824 )
(14.1002, 56.0845 )(14.5502, 56.3547 )
(15.0002, 56.7545 )(15.7502, 57.8277 )
(16.5002, 58.7015 )(17.2503, 59.0462 )
(18.0003, 59.8897 )(18.7503, 60.8694 )
(19.5003, 61.1351 )(20.2503, 61.7635 )
(21.0003, 62.7511 )(21.7503, 63.054 )
(23.4003, 64.536 )(25.2004, 65.4865 )
(27.9004, 67.2252 )(29.8504, 68.1363 )
(32.8505, 70.205 )(35.8505, 71.7478 )
(40.3506, 73.8998 )(45.6007, 76.2815 )
(49.0507, 77.366 )(51.9008, 78.6048 )
(55.9508, 80.4629 )(61.0509, 82.5349 )
(67.201, 86.0822 )(72.0011, 89.0991 )
(74.2511, 92.581 )(74.4011, 93.0957 )
(74.8511, 94.7219 )(75.3011, 95.7691 )
(75.4511, 96.6914 )(75.6011, 97.7027 )
(75.7511, 98.581 )(75.9011, 99.214 )
(76.0511, 99.6082 )(76.2011, 99.8232 )
(76.3511, 99.929 )(76.5011, 99.9763 )
(76.6511, 100.012 )}

\Curve{(0, 0 )
(0.15, 1.93018 )(0.300005, 3.85472 )
(0.60001, 7.42455 )(0.900015, 10.8412 )
(1.20002, 14.2613 )(1.65003, 18.9538 )
(2.10003, 23.2522 )(2.55004, 26.2444 )
(3.00005, 28.268 )(3.60005, 30.5924 )
(4.20007, 33.3682 )(4.80007, 35.7849 )
(5.40008, 37.2623 )(6.00009, 38.6239 )
(6.6001, 40.2894 )(7.2001, 41.8096 )
(7.80011, 42.8401 )(8.70013, 44.4223 )
(9.60014, 46.2443 )(10.5002, 47.4392 )
(11.4002, 48.643 )(12.3002, 49.9697 )
(13.2002, 50.7771 )(17.5503, 55.2579 )
(24.4504, 60.3186 )(30.3005, 63.4076 )
(35.1005, 65.6926 )(40.5006, 67.8322 )
(50.1007, 71.223 )(60.9009, 74.4752 )
(70.3511, 76.9155 )(80.2512, 79.3627 )
(85.3513, 80.6915 )(90.3013, 81.8378 )
(100.351, 84.3322 )(110.552, 87.0179 )
(117.002, 89.3739 )(120.902, 90.6993 )
(125.102, 96.1025 )(125.402, 97.3761 )
(125.702, 98.7432 )(126.002, 99.5631 )
(126.302, 99.8896 )(126.602, 100.012 )}

\Curve{(0, 0 )
(0.15, 2.02816 )(0.450005, 6.08333 )
(0.75001, 9.7455 )(1.05001, 12.9707 )
(1.35002, 15.8074 )(1.65003, 18.1093 )
(2.10003, 21.349 )(2.55004, 24.3592 )
(3.00005, 26.875 )(3.45005, 28.8141 )
(3.90006, 30.2928 )(4.65007, 32.9133 )
(5.40008, 35.6093 )(6.15009, 37.5484 )
(6.9001, 39.1385 )(7.65012, 40.8514 )
(8.70013, 42.4346 )(10.0502, 44.5845 )
(13.8002, 49.1171 )(16.8002, 51.8627 )
(20.8503, 55.1251 )(25.6504, 58.1791 )
(30.1504, 60.5698 )(39.9006, 64.6092 )
(45.6007, 66.4504 )(51.4508, 68.1475 )
(59.4009, 70.3558 )(68.551, 72.3963 )
(80.2512, 74.8604 )(90.1513, 76.7004 )
(98.2515, 78.2353 )(105.452, 79.3953 )
(111.752, 80.5068 )(124.802, 82.5518 )
(136.952, 84.553 )(145.952, 86.2152 )
(150.902, 87.1351 )(159.152, 88.7544 )
(167.252, 91.1678 )(175.803, 99.2219 )
(177.003, 100.012 )}

\Text(150,-70)[]{Fig.2}
\end{picture}
\end{center}

\newpage
\vspace*{2cm}

\begin{center}
\begin{picture}(225,120)(0,0)

\LongArrow(0,0)(225,0)
\LongArrow(0,0)(0,120)
\Text(225,-7)[rt]{$1/\ln t$}
\Text(-5,120)[tr]{$f_2$}

\SetWidth{0.3}
\Line(-2,27.27)(0,27.27)
\Text(-5,27.27)[r]{0.5}
\Line(-2,72.72)(0,72.72)
\Text(-5,72.72)[r]{0.6}

\Line(33.33,-2)(33.33,0)
\Text(33.33,-7)[t]{0.12}
\Line(100,-2)(100,0)
\Text(100,-7)[t]{0.14}
\Line(166.6,-2)(166.6,0)
\Text(166.6,-7)[t]{0.16}

\Line(19.6979, 0.84273 )(199.679, 97.7527 )

\Text(189.679, 92.3527 )[]{$\scriptscriptstyle \diamond$}
\Text(154.415, 72.1123 )[]{$\scriptscriptstyle \diamond$}
\Text(131.99, 61.4582  )[]{$\scriptscriptstyle \diamond$}
\Text(115.882, 52.1255 )[]{$\scriptscriptstyle \diamond$}
\Text(103.473, 45.33   )[]{$\scriptscriptstyle \diamond$}
\Text(93.4689, 40.6414 )[]{$\scriptscriptstyle \diamond$}
\Text(85.1408, 35.9727 )[]{$\scriptscriptstyle \diamond$}
\Text(78.0412, 31.9305 )[]{$\scriptscriptstyle \diamond$}
\Text(71.8769, 28.7309 )[]{$\scriptscriptstyle \diamond$}
\Text(66.446, 25.7423  )[]{$\scriptscriptstyle \diamond$}
\Text(61.6041, 23.5859 )[]{$\scriptscriptstyle \diamond$}
\Text(57.2446, 20.5364 )[]{$\scriptscriptstyle \diamond$}
\Text(53.2867, 18.5082 )[]{$\scriptscriptstyle \diamond$}
\Text(49.6678, 17.1168 )[]{$\scriptscriptstyle \diamond$}
\Text(46.3387, 15.1836 )[]{$\scriptscriptstyle \diamond$}
\Text(43.2595, 14.0323 )[]{$\scriptscriptstyle \diamond$}
\Text(40.3982, 11.6709 )[]{$\scriptscriptstyle \diamond$}
\Text(37.7281, 10.9423 )[]{$\scriptscriptstyle \diamond$}
\Text(35.2272, 9.45818 )[]{$\scriptscriptstyle \diamond$}
\Text(32.8768, 7.83955 )[]{$\scriptscriptstyle \diamond$}
\Text(30.6613, 6.33409 )[]{$\scriptscriptstyle \diamond$}
\Text(28.5671, 5.30182 )[]{$\scriptscriptstyle \diamond$}
\Text(26.5827, 4.36773 )[]{$\scriptscriptstyle \diamond$}
\Text(24.6979, 3.54273 )[]{$\scriptscriptstyle \diamond$}

\Text(112,-70)[]{Fig.3}
\end{picture}
\end{center}

\end{document}